# EUROPEAN HISTORICAL EVIDENCE OF THE SUPERNOVA OF AD 1054 SKY ABOVE EUROPE ON 4th JULY 1054

**Miroslav D. Filipović[1*], Jeffrey L. Payne[1], Thomas Jarrett[2], Nick F. H. Tothill[1], Dejan Urošević[3], Patrick J. Kavanagh[4], Giuseppe Longo[5], Evan J. Crawford[1], Jordan D. Collier[1,5] and Miro Ilić[6]**

[1] *Western Sydney University, School of Science, Locked Bag 1797, Penrith, NSW 2751, Australia*
[2] *University of Cape Town, The Inter-University Institute for Data Intensive Astronomy (IDIA), Department of Astronomy, Private Bag X3, Rondebosch 7701, South Africa*
[3] *University of Belgrade, Faculty of Mathematics, Department of Astronomy, Studentski trg 16, 11000 Belgrade, Republic of Serbia*
[4] *Dublin Institute for Advanced Studies, School of Cosmic Physics, 31 Fitzwilliam Place, Dublin 2, Ireland*
[5] *University Federico II, Department of Physics, Via Cinthia 6, I-80126 Napoli, Italy*
[6] *Trebinje Astronomical Association, Trebinje, Republic Srpska, Bosnia and Herzegovina*

**Abstract**

We investigate possible reasons for the absence of historical records of the supernova of 1054 in Europe. At the same time, we search for the new evidences as well. We establish that the previously acclaimed 'Arabic' records from ibn Butlan originate from Europe. As one of the most prominent scientists of the era, he was in Constantinople at the time of the supernova and actively participated in the medieval Church feud known as the Great Schism. Next, we reconstruct the European sky at the time of the event and find that the 'new star' (SN 1054) was in the west while the planet Venus was on the opposite side of the sky (in the east) with the Sun sited directly between these two equally bright objects, as documented in East-Asian records.

*Keywords:* History and Philosophy of Astronomy, symbols, supernovae: SN1054, ISM: Supernova Remnants, Christianity

## 1. Introduction

We are fortunate to live in a time when scientific methodology is well established, and discoveries can be questioned and re-examined. However, this was not the case in Europe during the eleventh century, the time of the Great Schism. Supernovae (SNe), such as SN 1006 and SN 1054 (whose remnant is known as M 1 or the Crab Nebula, which is shown in Figure 1), were not understood in the so-called "high middle ages period" [1, 2]. During this period, Constantine IX Monomachos 'The Lone Fighter' from the Macedonian dynasty was the Emperor of East Roman Empire, and he ruled from 11 June 1042 until 11 January 1055 (all dates in this paper refer to the Julian Calendar). In 1046-1047, he re-established what is arguably the oldest 'modern' university, the University of Constantinople - Pandidakterion, which was originally founded in 425 by Theodossius II [3, 4]. This was the most prestigious and advanced European scholarly centre at the time, with its two main schools: Philosophy and Law [5].

---
[*]E-mail: m.filipovic@westernsydney.edu.au

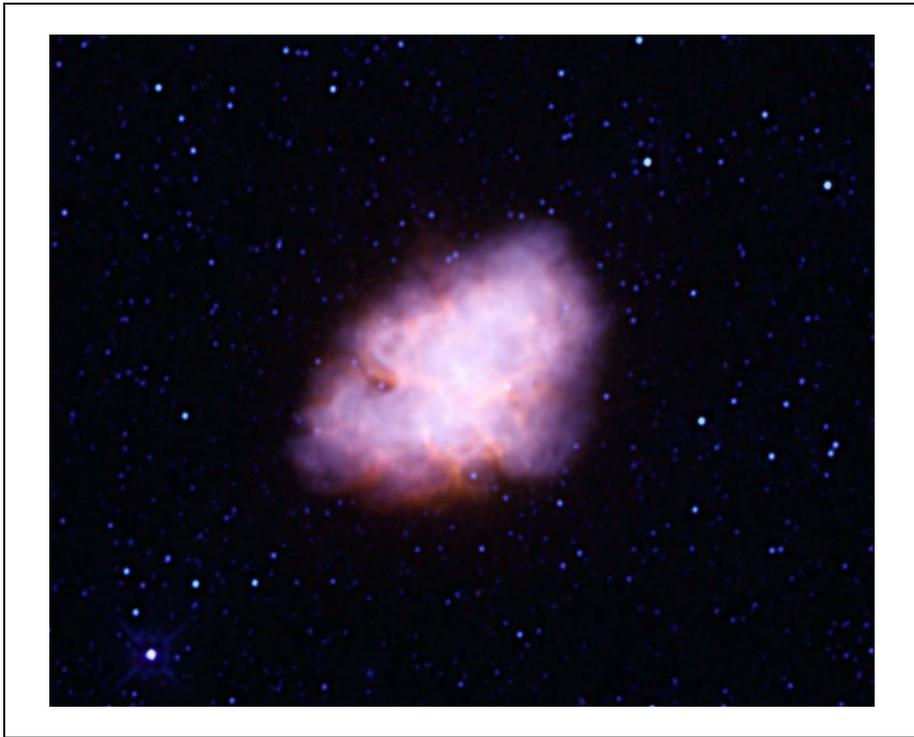

**Figure 1.** Wide-field Infrared Survey Explorer (WISE) image of the SN 1054 (a.k.a. Messier 1 (M1), NGC 1952, Sharpless 244 and the Crab Nebula), revealing the remnant of a great stellar explosion over 967 years ago.

At this time, Science in East Roman Empire (south-eastern Europe) was dominated by the middle-age doc- trine which in essence supports a flat Earth in the centre of Universe - a doctrine which the dogmatic Christian Church adopted as its official view of nature. This is all despite a number of experiments such as Hermannus Contractus (ca. AD 1013-1054; a.k.a. Hermann of Reichenau Abbey) who estimated the circumference of Earth with Eratosthenes' method as well as possibly the designer of one of the earlier medieval astrolabes. In this view, the planets moved in epicycles, to account for the Earth being at the centre of the Solar System, and there existed lunar and sub-lunar spheres that extended to the Moon, beyond which were the heavens where nothing ever changed.

According to this philosophy, all transient phenomena such as comets, novae and SNe therefore occurred within the lunar and sub-lunar regions. For the astronomer/astrologer of the time it was very dangerous to question such well-established 'laws', since they faced death for such 'blasphemy' - especially if it is coupled with the poisoned political affairs of the ruling Church clergy [6-9] and/or Horoscopy. As a testimony of this danger, Greek philosopher Psellos himself writes: "…but if he (Emperor) discovered men going so far as to utter blasphemies against the Lord Himself, he punished them by exile, or restricted their movements to a circumscribed area, or kept them in close confinement in prison, and he used to bind himself by secret oaths never to release them" [5, p. 69]. And then confesses himself: "…Because of my interest in horoscopes, I find myself inevitably subjected to troublesome inquiries about them. That I have applied myself to the Science in all its aspects I admit but at the same time none of these studies, forbidden by the leaders of the Church, has been put to improper use." [5, p. 76]

So, when SN 1054 occurred on 4$^{th}$ July 1054 in the northern constellation Taurus and was apparently visible in the daytime sky for 23 days, why was it not recorded in European history, when it was recorded in East-Asian [10-13] and Arabic records [14-16]? The Crab Nebula ($\alpha_{J2000} = 5^h34^m31.94^s$ and $\delta_{J2000} = +22°00'52.2"$, Figure 1) was first associated with the supernova of AD 1054 only in the XX$^{th}$ century [17, 18].

The most comprehensive reviews of the historical records of SN 1054 [19-22] concluded that there are no credible records of seeing SN 1054 explosion from Europe. They also rejected earlier claims of several possible European records [23-27] and suggested that knowledge of the Universe at the time was limited, which might explain why there were no records of SN 1054 from this part of the World (though see next paragraph). However, such knowledge was not required for one to see a bright stellar object appear and then fade away over time. It would certainly have been obvious to most people in broad daylight, but then again, maybe looking at the sky was not important or special - and if you did look then you were not able to accept or believe what your eyes saw!

One of the European records discussing observations of astronomical events is recorded in the Irish Annals, chronicles of medieval Irish history kept by monks [27]. A 1054 entry contained a possible reference to the SN 1054 event, a "round tower of fire at Ros Ela" visible "for five hours of the day". The main issue with this sighting is that the entry references the feast of Saint George, which fell on 24$^{th}$ April 1054, some 2-3 months earlier than the widely accepted and somewhat more precise East-Asian records. However, using topological and historical information of the region the authors hypothesised that the observer was located at a monastic complex near Ros Ela, and recorded the sighting of the bright object rising over Ros Ela in the early hours of a morning in July 1054, reconciling the time difference be- tween this and the East-Asian sightings. If the sighting was made in April, then it would have been an evening object nowhere near Ros Ela. After their careful examination of the interpolated language of the possibly incomplete entry, the location referenced in the entry and the location of SN 1054 in the sky, and the visibility time, they suggest that this may indeed constitute a European sighting of SN 1054. But, they caution that it is difficult to prove conclusively, though noting that if not referring to SN 1054, this was a remarkable coincidence. Stephenson and Green propose that a solar halo display may be responsible for this sighting but caution that "such a naive, materialistic interpretation may be inappropriate" [22]. It is very unlikely that the sighting would be a related to a solar halo. The Irish Annals contain only significant events in the medieval Irish skies such as solar/lunar eclipses, comets, blood moons, the aurora, etc. [27]. Solar halos and pillars are common phenomena and it seems unlikely they would be worthy of an Annals entry.

Several authors [28-30] have analysed various historical records and attempted to derive a light curve for SN 1054. However, [31-33] have all suggested that the Anasazi 'star map' in Chaco Canyon (New Mexico, USA) may be a North American record of SN 1054, but others [34-37] strongly refuted this interpretation arguing that the crescent-star-hand-Sun symbol is a sun-watching station. Finally, we note Armenian documents that could be linked to this SN explosion [38], but once again, the evidence is far from overwhelming.

Visible SN events are very rare in the Galaxy (Milky Way) and their occurrence is unpredictable. There are only 2-3 every hundred years in any typical spiral galaxy [39], which are generally not visible to the naked eye. Only the Milky Way Galaxy is accessible to astronomers and regular folk in the medieval times. At the present, not more than a dozen are recorded in our Galaxy over the past 2200 years. Therefore, they happen in average one every two hundred years within the few kpc of our Sun where they might be visible to the human eye [40] - a much longer time scale than the human lifespan. Hence, you are very lucky if you see one of these events during your lifetime. We are still waiting for a new and easily visible SN to appear in our Galaxy - the last one (Kepler's Supernova) was visible in AD 1604 as well as 'naked eye' visible SN1987A from the nearby galaxy Large Magellanic Cloud.

The absence of any indisputable European historical record of SN 1054 is intriguing as it was bright and would certainly have been a conspicuous object, especially during the Northern Hemisphere summer when people went outdoors regularly, gathered round large fires and spent the night in the open - thus having plenty of opportunity to observe the sky [41]. There can be little doubt that SN 1054 was seen and recognised as a 'new star' by most, if not all, cultures around the world. But how many created records of the event that survived through to modern times? The Aboriginal Australians did not have written records, and virtually nothing written survives from most of the African continent [34]. In contrast, the various Asian and Arab cultures had an effective bureaucracy that recorded new stars (both novae and SNe) in various dynastic annals that have survived through to the present day.

## 2. The historical context

It is essential to understand the historical context in which this astronomical event happened. Here, we investigate the political, cultural and scientific environment of the time around AD 1054 [21, p. 89].

To start with, this was a turbulent period when it came to the relationship between clergy of the Christian Church in Constantinople and in Rome. While the differences between the Patriarch in Constantinople and the Archbishop of Rome (the Pope) were on-going over several centuries, this particular period culminated in the 'Great Schism': the break in communion between what are now the Catholic Church and the Eastern Orthodox Church. The practices established then are still in place today, some ten centuries later [42].

It is certainly true that the large territory encompassing much of today's Southern Europe was difficult to rule from one location, as already discovered by a number of Roman Emperors during the II and III centuries AD. Various challenges were facing Christianity including internal rebellions and wars with neighbouring Barbarians, Russians, Seljuks, Pechenegs, Serbs, Croats, Armenians and probably the most dangerous of all, the continental Normans.

In Constantinople, power was shared by the Emperor and the Patriarch, while in Rome the Pope had everything under his control, including the military. In principle, the responsibility of an East Roman Emperor such as Constantine IX Monomachos (ca. AD 1000-1055) was to lead the state (he ruled from 1042 to 1055), but the influence of the Patriarch of Constantinople, Michael I Cerularius (ca. AD 1000-1059, r. 1043-1058), was such that in practice he could veto any decision made by the ruler. And indeed, this implies that these two contemporaries did not always enjoy a smooth working relationship but, note that the reliability and veracity of these Psellos memoirs are strongly questioned by [43]. On one side was Constantine IX with his very colourful social life, while on the other side was Cerularius, one of the most conservative and stubborn Patriarchs in the history of the Christian Church. Even though he was sick for much of his reign, Constantine IX was very popular with the people of Constantinople [44], but Cerularius did not enjoy similar support. Still, most of today's 'traditional' historians see Constantine IX as a very weak Emperor, despite the fact that he won at least two major wars (against the Russians and the Trache in 1043), re-established a university (in 1046), and treated people with greater dignity than ever before. The fact that after his death the State was weakened economically and militarily was surely a major sign of Constantine IX's success as an Emperor, and can be described as a "…capable, energetic, resourceful, and conscientious ruler, one of the best that empire ever had" [43, p. 181].

It is also worth mentioning that the Great Schism was, at the time, seen as a 'bad deal' for Rome and the Pope, as they were facing a new wave of imminent attacks by Normans from the north. These were the same Normans who invaded and conquered England twelve years later, in 1066. Indeed, Constantine IX's army was far superior to that of Pope Leo IX's (1002-1054, r.

1049-1054), as he was negotiating with the Pope and his administration to create a new alliance to defend Southern Italy from the Normans. Some would even suggest that the main East Roman Empire threat in 1054 was the Normans [43].

Most importantly, Leo IX died on 19$^{th}$ April 1054, three months before the appearance of SN 1054, and the next Pope, Victor II (1018-1057), was appointed on 13$^{th}$ April 1055 [42]. Therefore, during at least part of the Great Schism, Rome did not have a Pope.

The year of 1054 is well described in various letters from ibn Butlan (1001-1066), who was one of the most prominent Nestorian Christian physicians (doctors) and scientists of the time [14-16, 45]. He also witnessed a terrible contagion epidemic (plague) in Constantinople in the early summer of 1054, with about 14000 corpses in the city morgues, which clearly places him in Constantinople during the SN 1054 event [46]. Ibn Butlan was originally commissioned to work for the Patriarch of Constantinople and against Pope Leo IX [45]. He did this in collaboration with Nicetas Stethatos (1005-1085), a monk at the Stoudios and under the close supervision of Psellos himself [47]. Ibn Butlan's main job throughout March, April and May 1054 (just before the Great Schism) was to write an 'Essay on the Holy Eucharist' - a refutation of the Latin position on the controversial issue of whether or not one could use unleavened bread to celebrate the Eucharist [45]. This suggests that ibn Butlan was very close to all three: Emperor Constantine IX, Psellos and Patriarch Cerularius - which could have influenced or biased his judgement as an objective scientist. Once he was at a safe distance from Constantinople, in Cairo (Egypt) in 1056, he published his original notes about the SN 1054 event [15, 16]. At the time, Cairo was under Arab rule and hence ibn Butlan was no longer in fear of possible wrath from the Christian Church.

Meanwhile, the most influential scientist of the time was arguably the eunuch Michael Psellos (1017-1096), who was a Neo-Platonist and 'Consul (or President) of the Philosophers' at the newly re-established University of Constantinople. Apart from being a prominent philosopher, East Roman Empire writer, politician, historian and orator, he was one of the leaders of the higher society life in Constantinople. Psellos was undoubtedly a close follower of Plato's doctrine. From his book *Compendium Mathematicum*, which also includes music and Astronomy [48], we can see that his interest in astronomy was more about commenting on the works of others, rather than establishing new views - a fact rather common in those days. It is also suggested that Psellos was also someone who in today's terminology we would call a 'propaganda magnate' for the State and Emperor of the day [43]. Some even argue that in 1054 he was the main aide to the Emperor [47, 49].

At the time of the SN 1054 event, both ibn Butlan and Psellos were together in Constantinople witnessing one of the most prominent periods in the history of the Christian Church [50]. Technically, this immediately proves that European records of SN 1054 event do indeed exist. Interestingly, the date of the illegitimate (as there was no Pope) excommunication of the Patriarch of Constantinople Michael I Cerularius, 16[th] July 1054, corresponds to when SN 1054 may have reached its maximum brightness and allegedly was visible in the daytime [29]. Others also place both men (ibn Butlan and Psellos) quite close to Michael I Cerularius on the day of the Great Schism [20]. Ibn Butlan specifically described an obviously distressed and very worried Patriarch. However, it is important to emphasise that Psellos did not explicitly mention the Great Schism or SN 1054 in his memoirs, nor Comet 1P/Halley in 1066. This is somewhat surprising given that he single-handedly wrote the accusation [43, p. 221] in his eulogy at Patriarch Cerularius' funeral (on AD 21[st] January 1059), and Psellos clearly praised him for attacks on the false doctrine of the Latins [51]. One should not forget that Psellos was the highest-ranking politician of the time. Accordingly, immediately after Patriarch Cerularius' funeral he became the 'Prime Minister' of the East Roman Empire. Certainly, he knew how to navigate the complex political landscape of the mid XI century.

On the other side of the Christian Church, Cardinal Humbert of Silva Candida (1000/1015-1061) was most likely the real head of the leaderless Roman side of the Christian church. Cardinal Humbert was truly welcomed by Emperor Constantine IX [43], along with Frederick of Lorraine (1020-1058; later, Pope Stephen IX), but they were the ones who laid a papal bull of excommunication of the Patriarch on the high altar of the Cathedral of Hagia Sophia [42, p. 706]. This was all done despite the fact that Leo IX had died and the excommunication was therefore invalid [52]. The legate's mission in Constantinople started in mid-April 1054 and initially had the prime goal of securing Emperor Constantine IX's military support and alliance for the war against the ruthless Normans in southern Italy [43, 47]. And yet, they excommunicated the Patriarch, thus destabilising this needed partnership with Constantinople, which is very reconcile to understand strategically and politically.

It is certainly true that the Great Schism hardly influenced the life of ordinary people at the time, and some even examines it as a non-event [47]. Despite the 18-20[th] July 1054 riots on the streets of Constantinople against the attitude of the legates [43], the two parts of Christendom were not yet fully conscious of the great divide that existed between them. The Great Schism only had immediate repercussions for the Russian Church where there was competition between the Latin and East Roman Churches [47]. People on both sides still hoped that the misunderstandings would be cleared up without too much difficulty, as had occurred during similar previous crises. The dispute remained something that ordinary Christians in the East and the West were largely unaware of for several decades. It was the Crusades in the XIII[th] century that made the Great Schism definitive: they introduced a new spirit of hatred and bitterness, and brought the whole issue down to the (un)popular level [53]. Still, the Great Schism created a dangerous time for those at the top and in the leadership, enough to cover-up any bad portents (SN 1054).

## 3. Historical supernovae above Europe

One of the most conspicuous events in the sky is a supernova (SN) - the explosion that marks the violent end of a dying star. Because of Galactic obscuration (the gas-dust plane of the Galaxy), the limitation of the human eye, and especially distance, not all of these are visible from the Earth. Thus, the last one that was observed with the naked eye appeared more than four centuries ago (Kepler's SN, in AD 1604), while the youngest-known galactic Supernova Remnant (G1.9+0.3) is estimated to be around 120 years old [54-56]. The SN remnant Cassiopeia A is a recent (~320 years old) SN that was not recorded visually, since it is located in the Galactic plane, and would have been obscured [57]. Indeed, Type II SNe are preferentially seen in the Galactic plane and thus hidden, whereas Type Ia SNe can be anywhere in the sky. Type II supernovae result from the core collapse of massive stars preferentially formed in the Galactic plane from dense molecular clouds. Type Ia SNe represent a binary system detonation from mass transfer to white dwarfs from evolved stars or the merger of two white dwarfs.

In 1006, just 48 years before the appearance of SN 1054, another naked eye SN was visible from Europe. This was recorded in several European (Italy and Switzerland) chronicles [58-61] and in Arabic texts [62-64], but, interestingly, not in any extant East Roman Empire records. Several centuries after SN 1054, two more SNe appeared and were recorded by European astronomers, in 1572 (Tycho's SN) and in 1604 (Kepler's SN) [21, p. 89].

In 1572, 26-year old Danish astronomer Tycho Brahe (1546-1601) was determined to find out what really caused this bright object to appear in the sky [21, p. 89]. He employed his network of astronomical colleagues to find if there was an observable parallax, and the answer, which later changed our perception of the Universe, was 'no'. The conclusion was quite simple and not the first time it was suggested for variable objects - this SN must lie far beyond the Moon, amongst the fixed stars. Arabic text by Abū Maʿshar's (created seven centuries before Tycho Brahe) was examined [65] and shown how he came to the conclusion that a comet was behind Venus, i.e. supra-lunar and moving through the spheres. Abū Maʿshar's argument was incorrect, but it was cited by Tycho Brahe in his work on the new star of 1572. Clearly, changes that only occurred in the sub-lunar sphere, was incorrect. The next step for the astronomers of the time was rather challenging - they needed to tell the truth (or a very good story) to their very conservative Church leaders but still remain in good standing with the Canon.

So why did this discovery happen in 1572, but not some 518 years earlier, in 1054? Unfortunately, in 1054, European scientists were far more concerned about the astrological importance of such an occurrence than its astronomical importance. A SN therefore was a *harbinger of change* that was bad for the current rulers, including the clergy [66, 67], so it was easier and more politically expedient to ignore this bad omen rather than have it upset the established order. This is not to say that church members or clergy didn't know about or record these 'new' stars as is the case of Saint Gall and SN 1006 [60]. Official stances of the Church were often different from actions or beliefs of its members. An even more logical explanation as to why this change could occur in 1572 is because of Copernicus and his model of the Solar System, which competed with the old Ptolemaic system. Galileo and Kepler believed in the Copernican model, and were willing to carry out observations to test it. Notably, Tycho very much opposed Copernicus and in fact he proposed a system as an alternative to Copernicus' that did not violate the centrality of the Earth. However, this experimental or enlightened attitude did not exist in 1054. Also, one should not forget that the available astronomical instruments in the eleventh century were far inferior to those of the sixteenth century. This was of great importance when it came to measure parallaxes of nearby celestial objects and therefore constructing a picture of the 'new' Universe. Although one shouldn't forget that astronomy was quietly getting its moment in Europe during the late XI century with the knowledge of astrolabe - a device for measuring the position of heavenly bodies which was introduced to Europe from the Islamic world [68]. However, it was most likely not available to Constantinople scientist in 1054 [69].

On the other side of the world, Chinese and Japanese astronomers precisely recorded the SN 1054 event and dated its first sighting as a bright 'guest star' on 4$^{th}$ July 1054 [13]. They also noted that this star was visible in the daytime for 23 days and that at peak magnitude it was as bright as the planet Venus. They documented that the star faded some 642 days later, beyond recognition with the last sighting on 6$^{th}$ April 1056. Just how precise these East-Asian records are, has been the subject of numerous discussions [30, 37].

As stated above, ibn Butlan himself was one of the scholars from the University of Constantinople who was indeed living there at the time of the SN event [14, 16, 45]. So why did he not publish his observations in Greek (or Latin) while living in Constantinople? Was he simply writing in his native language (Arabic) or was he afraid of the political repercussions? Did the mighty Church clergy of the time orchestrate a massive cover up and deny the appearance of this new bright star? And if so, why? Did the Church clergy find it expedient to hide this dramatic portent from history? How brave would you have to be in a time of great political instability to observe, record and then report this rare and awesome event? One should not forget that people were superstitious in those days, and that omens from heaven were frightening. Also and as mentioned above, from early Greek times (V century B.C.), 'new stars' and comets were taken as atmospheric (sub Lunar) phenomena and not subject to Aristotelian immutability.

We set out to investigate this curious and notable omission from the historic records, and pose the question: did the scientists of the time, and indeed the Emperor himself, record this major event using unconventional and secret means to evade the Eye of the Church and if so - why?

## 4. The sky above Europe on 16th July 1054

We used the Stellarium Software Package to reconstruct the sky above Constantinople on 16th July 1054 (Figure 2) and found that Venus was in the east (in Leo; as well is at aphelion) and SN 1054 was in the west (in Taurus), with the Sun in Gemini directly between these two equally bright objects - as documented in the East-Asian records. The best position for both Venus and SN 1054 to be seen at the same time on the Constantinople sky would be after ~7:00 am when Venus would be seen as the morning star and SN 1054 at the opposite side in the west at ~30º from the zenith. Certainly, people would see bright new SN 1054 in the morning sky from ~1:30 am. Therefore, during the daytime both Venus and SN 1054 would be at the best on a border of visibility and Figure 2 picture would need to be reconstructed by astronomers at the time.

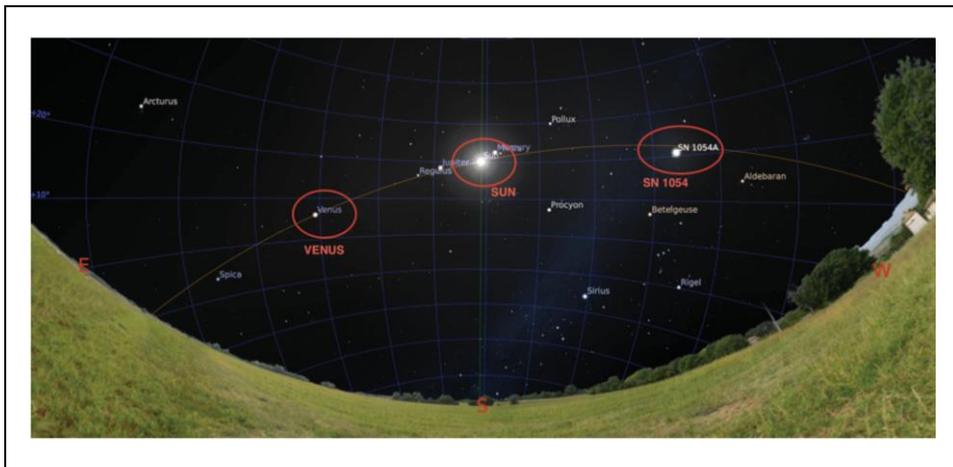

**Figure 2.** The sky above Constantinople on noon of 16th July 1054 (reconstructed using Stellarium software). Venus is in the east (at elevation of ~45º) and SN 1054 is in the west (at elevation of ~50º), with the Sun (at elevation of ~70º) sitting directly between the two equally bright objects.

The above scenario is what modern users of planetarium software will see, but in the summer of 1054 people would have seen the SN above the eastern horizon and about 12 hours later Venus above the horizon in the opposite direction. However, this is not mentioned in any of the European, Arab or Chinese sources that record the SN. Indeed, Schaefer (pers. comm., 2018) argues that there is little evidence that people thought about the visibility of planets or stars in the daytime (for a review of the visibility of stars in the daytime see [70-74]). However, we argue that any astronomer/astrologer at the time would not have difficulties to fully reconstruct this above scenario (from Figure 2) as long as two main objects (Venus and SN 1054) are as bright as they are reported.

One could also speculate that Constantine IX took advantage of the celestial event to affirm the centrality of the temporal authority between two Churches and the possible clash between him and Patriarch Cerularius. At the very time the star was visible, for the first time ever the Patriarch was not invited to attend the assembly of the ministers [43]. However, the Emperor and the Patriarch eventually made an alliance to counter Psellos' influence [43].

## 5. Concluding remarks

We survey the records of European history and culture from around 1054 to ask whether what we see is consistent with SN 1054 having been seen in the skies above Europe and having had an impact on the peoples who saw it. We have reviewed and analysed some of the factors that could account for the lack of European records of SN 1054. Perhaps the most plausible explanation for this would be the poor scientific knowledge (and interest) of celestial events such as supernovae at this time, and the added likelihood that there was a very subtle cover-up orchestrated by the Christian Churches as they both feel it is in their best interest to downplay the omen. Given the Church's stand on Astronomy/astrology, there would be a strong incentive not to report the occurrence of any event - including an obvious supernova - that would threaten the astronomical status quo and hence, God's Heaven as told by the Church Elders.

We do claim that the well-known 'Arabic' record of the SN 1054 sighting was indeed of European origin as the writer, ibn Butlan, was in Constantinople at the time he observed and recorded the supernova. Certainly, there are no other precise and indisputable European records of SN 1054 that are comparable to the ones from the East-Asian countries.

While there is no doubt that most (if not all) of the historical records around SN 1054 event suffer from various biases, we will present in part 2 and 3 of this series, a deeper dive into explanations for a lack of conventional records and the possibility some form of 'hidden' European record could relate directly to the SN 1054 event.

**Acknowledgment**

We thank Milan S. Dimitrijević, David Green, Ray Norris, Wayne Orchiston, Emilia Pazstor, Bradley Schaefer, Velibor Velović and the Western Sydney University Library team led by Linda Thornley for valuable help in reading and discussing various ideas regarding this study. We also thank Aleksandar Zorkić, Ain De Horta and Darren Maybour for technical help in preparing the figures shown here. We acknowledge that this paper made use of the Stellarium software (http://www.stellarium.org/).